\documentclass[10pt,a4paper,notitlepage]{article}
\usepackage[latin1]{inputenc}
\usepackage{verbatim,graphicx,amsmath,amsfonts,amssymb,mathrsfs}

\def\arraystretch{1.5}

\title{On the Geometry of Supersymmetric\\ Quantum Mechanical Systems\footnote{Supported by the Swedish Research Council}}
\author{
	D. Lundholm\footnote{dogge@math.kth.se}
}
\date{\scriptsize{Department of Mathematics, Royal Institute of Technology\\SE-10044 Stockholm, Sweden}}

\begin{document}

\maketitle

\begin{abstract}
	We consider some simple examples of supersymmetric quantum mechanical systems 
	and explore their possible geometric interpretation with the help
	of geometric aspects of real Clifford algebras.
	This leads to natural extensions of the considered systems to higher
	dimensions and more complicated potentials.
\end{abstract}

\vspace{10pt}

\section{Introduction}

	In the following, a supersymmetric system will mean a supersymmetric quantum
	mechanics (SUSYQM) according to the following definition\footnote{See 
	\cite{froehlich_et_al, combescure_et_al} and references therein for a discussion of possible definitions.}:
	On a complex separable Hilbert space acts a hamiltonian $H$,
	a number of supercharges $Q_{j=1,\ldots,\mathcal{N}}$, and a
	grading operator $K$ which splits the Hilbert space $\mathscr{H} = \mathscr{H}_b \oplus \mathscr{H}_f$
	into a bosonic and a fermionic sector.
	These operators are self-adjoint on their respective domains and satisfy the relations
	\begin{equation} \label{}
		\{Q_j,Q_k\} = 2\delta_{jk}H, \qquad K^2 = 1, \qquad \{Q_j,K\} = 0,
	\end{equation}
	where $\{A,B\} := AB + BA$ is the anticommutator.
	
	A classic example of a SUSYQM with a geometric interpretation is
	provided by the Dirac operator \cite{froehlich_et_al, kirchberg_et_al, cycon_et_al}.
	That is perhaps a rather uninteresting example in the flat space case, 
	where only local geometry is non-trivial, 
	but its extension to the setting of curved manifolds has led to new
	insights in global geometry and index theory.
	Here, we will focus on the local geometry of supersymmetric systems
	with Schr\"odinger-like hamiltonians.
	In particular, we are interested in hamiltonians of the form
	$H = H_B + H_F$, where the so-called bosonic part $H_B$ 
	is an ordinary Schr\"odinger operator
	and the fermionic part $H_F$ is a matrix- or algebra-valued multiplication operator.
	Since such operators involve a laplacian, their corresponding 
	supercharges will necessarily have to involve some form of Dirac operator.
	
	de Crombrugghe and Rittenberg \cite{crombrugghe_rittenberg}
	have carried out a rather general algebraic analysis of SUSYQM
	hamiltonians, but with focus on cases when the supercharges
	are linear in the Clifford generators. This is true for a single
	Dirac operator, but will apart from that generally not be
	the case in the examples we are considering.
	Furthermore,
	when studying the algebraic properties of supersymmetric systems 
	it is common to work with creation and annihilation operators 
	$a_j^{\phantom\dagger}, a_j^\dagger, c_k^{\phantom\dagger}, c_k^\dagger$
	and consider a Fock representation of these on a Hilbert space
	with a particle interpretation. 
	We will on the other hand stick to the alternative Schr\"odinger
	representation involving coordinates and momenta $x_j$, $p_{x_j}$ acting as
	multiplication and diffential operators on an $L^2$-space, 
	and Clifford generators $\boldsymbol{e}_{k1}$, $\boldsymbol{e}_{k2}$ acting in a representation
	of the corresponding Clifford algebra.
	We will emphasize the real geometry in the systems we consider
	and use it to `explain' the appearance of complex structures.
	This leads to the identification of additional structures, 
	properties and possible extensions of these systems
	which might not have been at all obvious from the conventional 
	complex formulation.
	We will also point out that it is possible to find a notion of
	supersymmetric system even in a purely real setting with
	no canonical complex structure.
	
	Apart from the purely mathematical interest in investigating the
	structure of these types of systems, one motivation from physics is that
	it seems worthwhile to explore the possibility of giving more
	complicated, but related, SUSYQM systems such as supersymmetric matrix models
	a more geometric interpretation.

\section{Geometric algebra}
	
	In order to appreciate the geometric interpretation of the
	systems which we consider it is helpful to have available some tools and
	notions from geometric algebra, i.e. Clifford algebra
	with emphasis on the geometry of the underlying, usually real, vector space.
	For a more complete introduction to the geometric aspects of Clifford
	algebra, see e.g. \cite{doran_lasenby} or \cite{lundholm}.
	
	Given a real vector space $V$ of dimension $d$ with a
	non-degenerate bilinear form $\boldsymbol{a} \cdot \boldsymbol{b}$ 
	(e.g. an inner product, Minkowski metric, etc.)
	there is a naturally associated Clifford algebra or \emph{geometric algebra} $\mathcal{G}(V)$
	in the following way. For an orthonormal basis $E = \{\boldsymbol{e}_1,\ldots,\boldsymbol{e}_d\}$ of $V$
	we let $\mathcal{G}(V)$ (or just $\mathcal{G}$) 
	denote the free associative algebra generated by $E$
	with the relations $\boldsymbol{e}_i^2 = \boldsymbol{e}_i \cdot \boldsymbol{e}_i = \pm 1$ 
	and $\boldsymbol{e}_i \boldsymbol{e}_j = -\boldsymbol{e}_j \boldsymbol{e}_i$, $i \neq j$.
	Hence, normalized vectors square to unity, orthogonal vectors anticommute, and
	\begin{equation} \label{}
		\mathcal{G}(V) = \textrm{Span}_\mathbb{R} \{1, \boldsymbol{e}_i, \boldsymbol{e}_i\boldsymbol{e}_j, \ldots, \boldsymbol{e}_1\boldsymbol{e}_2\ldots \boldsymbol{e}_d\}_{i<j<\ldots}.
	\end{equation}
	As vector spaces, $\mathcal{G}(V)$ is isomorphic to the graded
	exterior algebra $\bigwedge^* V$, and there is a corresponding
	exterior product $\wedge$ in $\mathcal{G}(V)$ with respect to which
	this isomorphism extends to the level of algebras.
	Hence we can identify these spaces.
	We let $\mathcal{G}^\pm$ denote the subspaces of even/odd grades in $\mathcal{G}$.
	The highest grade element $I := \boldsymbol{e}_1\boldsymbol{e}_2 \ldots \boldsymbol{e}_d = \boldsymbol{e}_1 \wedge \boldsymbol{e}_2 \wedge \cdots \wedge \boldsymbol{e}_d$ 
	determines an orientation for $V$ and is called the pseudoscalar.
	An arbitrary element of $\mathcal{G}$ is called a multivector.
	
	The orthogonal groups have spin representations embedded in $\mathcal{G}$ as
	can be seen by action of Clifford multiplication on $V \subseteq \mathcal{G}$.
	For example, a reflection along (i.e. in the hyperplane orthogonal to) 
	a unit vector $\boldsymbol{n} \in S^{d-1} \subseteq \mathbb{R}^d$ 
	acting on a vector $\boldsymbol{v} \in \mathbb{R}^d$
	can be written $\boldsymbol{v} \mapsto -\boldsymbol{nvn}$. 
	A rotation (being a composition of an even number of reflections)
	has the \emph{spin} or \emph{rotor} representation 
	$\boldsymbol{v} \mapsto R^\dagger \boldsymbol{v} R$, where $R = \boldsymbol{n}_1\boldsymbol{n}_2\ldots \boldsymbol{n}_{2k}$
	is a product of unit vectors and the dagger denotes reversion of the order of any Clifford
	product. The Dirac operator 
	\begin{equation} \label{dirac_op}
		\nabla := \sum_{j=1}^d \boldsymbol{e}_j \frac{\partial}{\partial x_j}
	\end{equation}
	in $V$ (flat space) 
	is of course also a natural object, combining the properties of differentiation with
	the properties of the Clifford product.
	
	A spinor is an element of a (often irreducible) representation of a spin group
	and as such can also be viewed as an element of the geometric algebra
	itself. Namely, any irreducible spin representation can be constructed by
	letting the spin group $\textrm{Spin} \subseteq \mathcal{G}$ act on an ideal of $\mathcal{G}$
	(or some other invariant subspace such as $\mathcal{G}^+$)
	by left multiplication. In this way, also any irreducible spinor bundle $S(M)$ 
	on a manifold $M$
	with associated principal spin bundle $P_{\textrm{Spin}}(M)$
	can be viewed as coming from the Clifford bundle $C\ell(M)$ associated with the tangent space,
	but with left multiplication $\ell$ instead of adjoint action Ad.
	In other words,
	\begin{equation} \label{}
	\begin{array}{rcl}
		C\ell(M) &=& P_{\textrm{Spin}}(M) \times_{Ad} \mathcal{G}(V), \\
		S(M) &\subseteq& P_{\textrm{Spin}}(M) \times_\ell \mathcal{G}(V),
	\end{array}
	\end{equation}
	where $V \cong T_pM$ for any $p \in M$. 
	(These bundle constructions will not be needed in the following; 
	see e.g. Example 3.7, Chapter II in \cite{lawson_michelsohn}
	for details.)
	
	For euclidean spaces the associated geometric algebra has a
	positive definite scalar product\footnote{Symmetry follows from 
	$\langle A^\dagger \rangle_0 = \langle A \rangle_0$
	and $\langle AB \rangle_0 = \langle BA \rangle_0$.}
	$\langle \cdot, \cdot \rangle_\mathcal{G}\!: \mathcal{G} \times \mathcal{G} \to \mathbb{R}$ 
	given by
	\begin{equation} \label{ga_scalar_prod}
		\langle A, B \rangle_\mathcal{G} := \langle A^\dagger B \rangle_0,
	\end{equation}
	where $\langle \cdot \rangle_p$ denotes projection onto the 
	grade-$p$ part of the expression. This scalar product is inherited by the
	chosen spinor space $\mathcal{S} \subseteq \mathcal{G}$ and can also be used to
	construct a hermitian inner product on $\mathcal{S}$ as will be 
	thoroughly explained in the examples.

\section{The Dirac operator}

	As a first example of a SUSYQM system in this formalism we consider the Dirac operator
	on $\mathbb{R}^d$. 
	A spinor on $\mathbb{R}^d$ should be an element of a representation space $\mathcal{S}$
	for $\textrm{Spin}(d)$. 
	According to the above we can choose $\mathcal{S} = \mathcal{G}(\mathbb{R}^d)$ or
	let $\mathcal{S}$ be some other invariant subspace of $\mathcal{G}$.
	Sometimes this space comes equipped with a natural complex structure,
	such as for $\mathcal{G}^+(\mathbb{R}^2) \cong \mathbb{C}$ 
	and $\mathcal{G}(\mathbb{R}^3) \cong \mathbb{C}^{2 \times 2}$, 
	in which case the real representation extends to a complex one.
	Sometimes we can nevertheless find a complex representation by
	e.g. acting on a spinor from the right with an element of $\mathcal{G}$
	squaring to $-1$
	(assuming then that the chosen spinor space is right-invariant under that action).
	For example, in dimensions such that $I^2 = -1$ (equivalently $I^\dagger = -I$)
	we can consider the \emph{real} Hilbert space 
	\begin{equation} \label{}
		\mathscr{H} := L^2(\mathbb{R}^d; \mathcal{G}(\mathbb{R}^d))
		\cong L^2(\mathbb{R}^d) \otimes \mathcal{G}(\mathbb{R}^d)
	\end{equation}
	and define an operator $\hat{\imath}$ on $\mathscr{H}$ by $\hat{\imath}(\Psi) := \Psi I$. 
	Since left-action commutes
	with right-action this will behave just like the ordinary complex imaginary $i$,
	thus providing a complex structure\footnote{Note that $\hat{\imath}$ not only
	squares to $-1$ but is also orthogonal 
	w.r.t \eqref{ga_scalar_prod}.} for the Hilbert space.
	The question of choice of hermitian inner product on $\mathscr{H}$ 
	will be addressed in the next example.
	
	A hermitian supercharge operator corresponding to
	the Dirac operator is e.g. given by $Q := -\hat{\imath}\nabla$.
	This squares to the positive
	self-adjoint hamiltonian $H := -\Delta$, i.e. minus the laplacian on $\mathbb{R}^d$.
	For \emph{even} dimensions such that $I^2 = -1$, i.e. $d = 2,6,10,\ldots$,
	there is also a natural grading operator given by $K := -\hat{\imath}I$.
	This anticommutes with the Dirac operator (a grade 1 object)
	and has value $+1$ on $\mathcal{G}^+$ and $-1$ on $\mathcal{G}^-$,
	hence it induces a boson/fermion grading of the Hilbert space
	into even and odd elements of $\mathcal{G}$:
	$\mathscr{H}_{\pm} := L^2(\mathbb{R}^d; \mathcal{G}^\pm(\mathbb{R}^d))$.
	The space $\mathscr{H} = \mathscr{H}_+ \oplus \mathscr{H}_-$
	together with the operators $H, K, Q$
	form an $\mathcal{N}=1$ supersymmetric system.
	(As pointed out in \cite{combescure_et_al}, we actually have $\mathcal{N}=2$.)
	
	Note that we could also consider the anti-hermitian supercharge $\tilde{Q} := \hat{\imath}Q = \nabla$,
	i.e. the actual Dirac operator, and regard this as a `supersymmetric system'
	$(\mathscr{H}, H, P, \tilde{Q})$
	with $H = -\tilde{Q}^2$ and $\{P, \tilde{Q}\} = 0$, 
	where $P\!: \Psi(\boldsymbol{x}) \mapsto \Psi(-\boldsymbol{x})$ is the parity operator.
	Another option is to define $K$ by $K(\Psi) := \Psi^\star$,
	where $A^\star := \pm A$, for $A \in \mathcal{G}^\pm$, is the grade involution of $\mathcal{G}$.
	This leaves out the need for 
	an operator $\hat{\imath}$
	completely and puts no restriction on the dimension $d$.
	The Hilbert space would in general be purely real 
	(i.e. without any complex structure) 
	in this canonical construction, however, 
	unless we formally complexify $\mathcal{G}$ or the spinor subspace $\mathcal{S}$.
	
	As usual, we can also extend the Dirac operator to be covariant under local 
	U(1) gauge transformations w.r.t. the complex structure,
	by introducing a connection vector field $\boldsymbol{A} = \sum_j A_j \boldsymbol{e}_j$.
	The supercharge then becomes 
	\begin{equation} \label{}
		Q = -\hat{\imath}\sum_j \boldsymbol{e}_j(\partial_j - \hat{\imath}A_j) = -\hat{\imath}\nabla - \boldsymbol{A}.
	\end{equation}
	The hamiltonian is now
	\begin{equation} \label{}
	\begin{array}{rl}
		H &= (-\hat{\imath}\nabla - \boldsymbol{A})^2 \\
		&= \sum_j \boldsymbol{e}_j\boldsymbol{e}_j (\hat{\imath}\partial_j + A_j)^2 + \sum_{j \neq k} \boldsymbol{e}_j\boldsymbol{e}_k \frac{1}{2} [\hat{\imath}\partial_j + A_j, \hat{\imath}\partial_k + A_k] \\
		&= (-\hat{\imath}\nabla - \boldsymbol{A})^2|_\textrm{scalar} + \hat{\imath}F,
	\end{array}
	\end{equation}
	where $(-\hat{\imath}\nabla - \boldsymbol{A})^2|_\textrm{scalar} := \sum_j (-\hat{\imath}\partial_j - A_j)^2 \geq 0$ 
	is the usual `scalar representation' of the operator $(-\hat{\imath}\nabla - \boldsymbol{A})^2$,
	and $F := \nabla \wedge \boldsymbol{A} = \sum_{j<k} \boldsymbol{e}_j\boldsymbol{e}_k (\partial_j A_k - \partial_k A_j)$ 
	is the field strength 2-form corresponding to $\boldsymbol{A}$ which appears as an
	extra term in the spinor representation.
	With $d = 3$, $\mathcal{S} = \mathcal{G}^+(\mathbb{R}^3)$, and the
	identification\footnote{The representation of $\mathcal{G}$ chosen in this case 
	is not purely left-action on $\mathcal{S}$, but a combination of left- and right-action
	in order to keep $\mathcal{S}$ invariant. $\textrm{Spin}(3)$ will 
	still be represented by left-action, however.} 
	of $\mathcal{S} \cong \mathbb{C}^2$ for spinors
	in three-dimensional space e.g. considered in \cite{doran_lasenby},
	this $H$ becomes the supersymmetric Pauli hamiltonian for a spin-$\frac{1}{2}$ particle
	in a three-dimensional magnetic field (see e.g. \cite{combescure_et_al}, \cite{cycon_et_al}, or \cite{froehlich_et_al}).

\section{Supermembrane toy model}

	Our main example is often considered as a simple toy model for a super\-membrane.
	This is because it shares many features with a more complicated SUSYQM
	matrix model which arises from a certain regularization of a su\-permembrane,
	or equivalently from dimensional reduction of supersymmetric Yang-Mills theory.

\subsection{Formulation of the model}
	
	In a common representation, 
	this system is described by the hamiltonian
	\begin{equation} \label{}
		H := (p_x^2 + p_y^2 + x^2y^2)1 + x\sigma_3 + y\sigma_1,
	\end{equation}
	acting on $\mathbb{C}^2$-valued wavefunctions 
	$\psi \in \mathscr{H}' := L^2(\mathbb{R}^2; \mathbb{C}^2)$.
	The $\sigma_j$ are Pauli matrices satisfying $\sigma_1\sigma_2\sigma_3 = i$.
	The corresponding supercharge is
	\begin{equation} \label{}
		Q := p_x\sigma_3 - p_y\sigma_1 - xy\sigma_2,
	\end{equation}
	and there is an additional discrete symmetry of the hamiltonian
	expressed by the `reflection' operator
	\begin{equation} \label{}
		(P\psi)(x,y) := \frac{1}{\sqrt{2}}(\sigma_1 + \sigma_3)\psi(y,x).
	\end{equation}
	The system $(\mathscr{H}', H, P, Q)$ exhibits 
	supersymmetry:
	\begin{equation} \label{}
		H = Q^2, \qquad P^2 = 1, \qquad \{Q,P\} = 0.
	\end{equation}

	The main property of this model which makes it interesting to study 
	both from a physics and a functional analysis perspective 
	is the form of the potential term $V := x^2y^2$ (cp. \cite{hoppe, deWit_luscher_nicolai}). 
	It is non-negative, but has valleys of zero energy extending
	to infinity along the coordinate axes. Because of this,
	a classical (scalar) particle described by the bosonic hamiltonian $H_B := p_x^2 + p_y^2 + V$
	has the possibility to escape to infinity, i.e. is unconfined. 
	However, the corresponding quantum mechanical system \emph{is} confined, 
	because the operator $H_B$ has a discrete spectrum.
	This can be understood intuitively by the fact that the valleys become steeper
	and narrower towards infinity which forces a wave packet to localize, thus increasing
	its kinetic energy due to the uncertainty principle.
	The discreteness of the spectrum of $H_B$ was first proved in a more
	general context in \cite{simon} but, as noted there, can also easily be seen from the operator
	inequality (in quadratic form sense)
	\begin{equation} \label{}
		H_B = \frac{1}{2}\left( p_x^2 + p_y^2 + (p_x^2 + y^2x^2) + (p_y^2 + x^2y^2) \right) 
		\geq \frac{1}{2}\left( p_x^2 + p_y^2 + |y| + |x| \right) \gg 0,
	\end{equation}
	i.e. the spectrum is discrete and bounded away from zero.
	On the other hand, it was proved in \cite{deWit_luscher_nicolai} that when one
	considers the supersymmetric hamiltonian $H$, the negative contribution from the
	fermionic term $H_F := x\sigma_3 + y\sigma_1$ precisely cancels the minimal bosonic
	energy in the valleys, resulting in $H$ having a continuous spectrum on the
	whole positive real axis (the bound below by zero follows from the supersymmetry algebra). 
	That does not rule out the existence of normalizable eigenstates, however,
	but it was shown in \cite{graf_hasler_hoppe} that zero 
	(which is the most interesting point in the spectrum, both with supersymmetry in mind,
	and from the relation of this model to more complicated SUSYQM systems)
	is \emph{not} an eigenvalue.
	The question of existence of embedded nonzero eigenvalues in the continuous
	spectrum is still open, although a positive answer has been suggested by numerical methods
	\cite{korcyl}.
	There is also a certain interest in understanding the generalized (distributional)
	zero energy eigenfunction(s) beyond smoothness (i.e. elliptic regularity \cite{browder})
	and decay rate towards infinity along a valley \cite{many_author_paper}.

\subsection{Geometric formulation}

	Our aim is to find a geometric interpretation of all aspects of this system
	by reformulating it in terms of geometric algebra. 
	We do this reformulation in some detail in order to
	eventually make an effortless transition to a more general setting.
	The form of the resulting model will also serve to `guide' us through a certain coordinate
	thansformation which could lead to a better understanding of
	possible eigenfunctions.
	Some of the geometric features of the model have been considered in \cite{korcyl},
	but in that case as a three-dimensional spin system in the plane.
	Here we will focus on the geometry of the plane itself.
	
	We start by using the representation $p_j = -i\partial_j$ and denote 
	$\gamma_1 := \sigma_3$, $\gamma_2 := -\sigma_1$, and 
	$\gamma_3 := -\sigma_2 = -i\gamma_1\gamma_2$.
	Then the hamiltonian becomes
	\begin{equation} \label{gamma_h}
		H = -\Delta + x^2y^2 + x\gamma_1 - y\gamma_2,
	\end{equation}
	and we will 
	instead of $Q$
	consider the anti-hermitian supercharge
	\begin{equation} \label{gamma_q_tilde}
		\tilde{Q} := iQ = \gamma_1\partial_x + \gamma_2\partial_y + xy\gamma_1\gamma_2,
	\end{equation}
	so that $H = -\tilde{Q}^2$. Furthermore, in this notation
	\begin{equation} \label{gamma_p}
		(P\psi)(x,y) = \frac{1}{\sqrt{2}}(\gamma_1 - \gamma_2)\psi(y,x).
	\end{equation}
	For concreteness, we choose the representation of $\gamma_j$ (or equivalently $\sigma_j$)
	such that
	\begin{equation} \label{gamma_rep}
		\def\arraystretch{1.0}
		\gamma_1 = \left[
		{
		\setlength\arraycolsep{3pt}
		\begin{array}{cc}
			0 & 1 \\ 1 & 0
		\end{array}
		}
		\right], \quad
		\gamma_2 = \left[
		\begin{array}{cc}
			0 & -i \\ i & 0
		\end{array}
		\right], \quad
		\gamma_3 = \left[
		\begin{array}{cc}
			1 & 0 \\ 0 & -1
		\end{array}
		\right].
	\end{equation}

	Note that neither $i$ nor $\gamma_3$ enter in the expressions \eqref{gamma_h}-\eqref{gamma_p}
	for the operators $H$, $\tilde{Q}$ and $P$. Remaining are only the coordinate
	multiplication operators $x,y$, 
	their corresponding partial differentiations $\partial_x, \partial_y$, 
	and the matrices $\gamma_1$ and $\gamma_2$ which generate a (matrix representation of a)
	Clifford algebra over two dimensions.
	Since the bosonic configuration space of this model is $\mathbb{R}^2$
	it is actually natural to consider the geometric algebra generated by the orthonormal
	standard basis $\{\boldsymbol{e}_x, \boldsymbol{e}_y\}$ in $\mathbb{R}^2$ 
	corresponding to the $(x,y)$-coordinate system, i.e.
	$\mathcal{G}(\mathbb{R}^2) = \textrm{Span}_\mathbb{R} \{1,\boldsymbol{e}_x,\boldsymbol{e}_y,\boldsymbol{e}_x\boldsymbol{e}_y\}$.
	Note that $I = \boldsymbol{e}_x\boldsymbol{e}_y$ and $I^2 = -1$, 
	implying $\mathcal{G}^+ \cong \mathbb{C}$.
	With the identification $\gamma_1 \leftrightarrow \boldsymbol{e}_x$ 
	and $\gamma_2 \leftrightarrow \boldsymbol{e}_y$,
	we then have
	\begin{equation} \label{}
		H = -\Delta + x^2y^2 + x\boldsymbol{e}_x - y\boldsymbol{e}_y 
	\end{equation}
	and
	\begin{equation} \label{}
		\tilde{Q} = \boldsymbol{e}_x\partial_x + \boldsymbol{e}_y\partial_y + xy\boldsymbol{e}_x\boldsymbol{e}_y = \nabla + xyI.
	\end{equation}
	These are now acting by left multiplication on the real Hilbert space 
	\begin{equation} \label{}
		\mathscr{H} := L^2(\mathbb{R}^2;\mathcal{G}(\mathbb{R}^2)).
	\end{equation}
	Hence, an arbitrary wavefunction $\Psi \in \mathscr{H}$
	can be written
	\begin{equation} \label{psi_ga}
	\begin{array}{rl}
		\Psi &= \psi_\emptyset + \psi_x \boldsymbol{e}_x + \psi_y \boldsymbol{e}_y + \psi_{xy} I \\
		&= (\psi_\emptyset + \psi_{xy}I) + \boldsymbol{e}_x(\psi_x + \psi_yI),
	\end{array}
	\end{equation}
	where $\psi_\emptyset,\psi_x,\psi_y,\psi_{xy}$ are real-valued functions.
	This makes the correspondence
	\begin{equation} \label{corresp_psi}
		\def\arraystretch{1.0}
		\psi = 
		\left[
		\begin{array}{c}
			\psi_1 \\ \psi_2
		\end{array}
		\right]
		=
		\left[
		\begin{array}{c}
			\psi_\emptyset + \psi_{xy}i \\ \psi_x + \psi_yi
		\end{array}
		\right] \leftrightarrow \Psi
	\end{equation}
	between $\mathscr{H}'$ and $\mathscr{H}$ explicit. 
	We find that the identification in \eqref{psi_ga} and \eqref{corresp_psi} 
	really is the desired one in this representation of $\gamma_j$ since e.g.
	\begin{equation} \label{corresp_ex}
		\def\arraystretch{1.0}
		\gamma_1 \psi =
		\left[
		\begin{array}{c}
			\psi_x + \psi_yi \\ \psi_\emptyset + \psi_{xy}i
		\end{array}
		\right]
		\quad \leftrightarrow \quad \boldsymbol{e}_x \Psi = (\psi_x + \psi_yI) + \boldsymbol{e}_x(\psi_\emptyset + \psi_{xy}I)
	\end{equation}
	and
	\begin{equation} \label{corresp_ey}
		\def\arraystretch{1.0}
		\gamma_2 \psi =
		\left[
		\begin{array}{c}
			\psi_y - \psi_xi \\ -\psi_{xy} + \psi_\emptyset i
		\end{array}
		\right]
		\quad \leftrightarrow \quad \boldsymbol{e}_y \Psi = (\psi_y - \psi_xI) + \boldsymbol{e}_x(-\psi_{xy} + \psi_\emptyset I).
	\end{equation}
	Furthermore, 
	even though we have not had use for it yet, in this two-dimensional setting
	there is also the natural operator $\hat{\imath}$
	taking the role as imaginary unit:
	\begin{equation} \label{corresp_i}
		i\psi  \quad \leftrightarrow \quad \hat{\imath} \Psi = \Psi I.
	\end{equation}
	Note that $\gamma_3 \leftrightarrow -\hat{\imath}I$ coincides with the
	grading operator $K$ defined in the previous section.
	This is consistent with the fact that $\psi_1$ and $\psi_2$ represent
	even and odd elements of $\mathcal{G}$, respectively.
	However, in this system, $K$ is no longer a symmetry of the hamiltonian $H$.
	
	Let us now turn to the issue of an inner product on $\mathscr{H}$.
	We have so far considered $\mathscr{H}$ as a real Hilbert space
	with an inner product inherited from the real scalar product 
	\eqref{ga_scalar_prod} on $\mathcal{G}$, i.e.
	\begin{equation} \label{real_inner_prod}
		\langle \Phi, \Psi \rangle_\mathbb{R} 
		:= \int_{\mathbb{R}^2} \langle \Phi(\boldsymbol{x}), \Psi(\boldsymbol{x}) \rangle_\mathcal{G} dxdy
		 = \int_{\mathbb{R}^2} \langle \Phi(\boldsymbol{x})^\dagger \Psi(\boldsymbol{x}) \rangle_0 dxdy.
	\end{equation}
	However, using the complex structure induced by the operator $\hat{\imath}$
	and a standard technique, we can construct a hermitian inner product
	from the real one:
	\begin{equation} \label{complex_inner_prod}
		\langle \Phi, \Psi \rangle_\mathbb{C} := \langle \Phi, \Psi \rangle_\mathbb{R} - i\langle \Phi, \hat{\imath}\Psi \rangle_\mathbb{R},
	\end{equation}
	so that e.g. $\langle \Phi, \hat{\imath}\Psi \rangle_\mathbb{C} = i\langle \Phi, \Psi \rangle_\mathbb{C}$.
	Note that for $\Phi, \Psi \in \mathcal{G}$
	\begin{equation} \label{}
		\langle \Phi, \Psi \rangle_\mathcal{G} = \langle \Phi^\dagger \Psi \rangle_0
		= \phi_\emptyset\psi_\emptyset + \phi_{xy}\psi_{xy} + \phi_x\psi_x + \phi_y\psi_y
		= \textrm{Re}\ \phi^\dagger \psi,
	\end{equation}
	where $\phi^\dagger = [\phi_1^*, \phi_2^*] = [\phi_\emptyset - \phi_{xy}i, \phi_x - \phi_yi]$,
	and hence
	\begin{equation} \label{}
		\langle \Phi, \Psi \rangle_\mathcal{G} - i \langle \Phi, \hat{\imath}\Psi \rangle_\mathcal{G} 
		= \textrm{Re}\ \phi^\dagger \psi - i \textrm{Re}\ \phi^\dagger i\psi 
		= \textrm{Re}\ \phi^\dagger \psi + i \textrm{Im}\ \phi^\dagger \psi 
		= \phi^\dagger \psi.
	\end{equation}
	Thus, we have actually constructed the standard inner product on 
	the complex Hilbert space $\mathscr{H}' = L^2(\mathbb{R}^2;\mathbb{C}^2)$,
	\begin{equation} \label{}
		\langle \Phi, \Psi \rangle_\mathbb{C} = \int_{\mathbb{R}^2} \phi^\dagger \psi.
	\end{equation}
	Now, since $\mathbb{C} \cong \mathcal{G}^+$ we might as well represent the
	value of this inner product in $\mathcal{G}^+$ itself by instead of \eqref{complex_inner_prod} taking
	\begin{equation} \label{}
		\langle \Phi, \Psi \rangle := \langle \Phi, \Psi \rangle_\mathbb{R} - \langle \Phi, \hat{\imath}\Psi \rangle_\mathbb{R} I
		= \int_{\mathbb{R}^2} \left( \langle \Phi^\dagger \Psi \rangle_0 - \langle \Phi^\dagger \Psi I \rangle_0 I \right).
	\end{equation}
	This is in fact a convenient expression since for any $A \in \mathcal{G}(\mathbb{R}^2)$
	\begin{equation} \label{}
		\langle A \rangle_0 + \langle A I^\dagger \rangle_0 I = \langle A \rangle_0 + \langle A \rangle_2 = \langle A \rangle_+,
	\end{equation}
	i.e. the projection onto the even part of $A$.
	One can also directly verify that
	\begin{equation} \label{}
		\langle \Phi^\dagger \Psi \rangle_+ 
		= (\phi_\emptyset - \phi_{xy}I)(\psi_\emptyset + \psi_{xy}I) + (\phi_x - \phi_yI)(\psi_x + \psi_yI) 
		\leftrightarrow \phi_1^* \psi_1 + \phi_2^* \psi_2 = \phi^\dagger \psi.
	\end{equation}
	The $\mathcal{G}^+$-valued inner product on $\mathscr{H}$ is therefore given by
	\begin{equation} \label{inner_prod}
		\langle \Phi, \Psi \rangle = \int_{\mathbb{R}^2} \langle \Phi(\boldsymbol{x})^\dagger \Psi(\boldsymbol{x}) \rangle_+ dxdy,
	\end{equation}
	and hence the expectation value of a self-adjoint operator 
	$\mathcal{O}\!: \mathscr{H} \to \mathscr{H}$ in the state $\Psi$:
	\begin{equation} \label{}
		\langle \mathcal{O} \rangle_{\Psi} := \langle \Psi, \mathcal{O} \Psi \rangle 
		= \int_{\mathbb{R}^2} \langle \Psi^\dagger \mathcal{O} \Psi \rangle_+.
	\end{equation}
	This has a clear geometric interpretation; e.g. for multivector-valued multiplication operators 
	$\Gamma \in L^\infty(\mathbb{R}^2;\mathcal{G}(\mathbb{R}^2))$ 
	and even wavefunctions $\Psi = \rho R \in \mathscr{H}_+$, where
	$\rho(\boldsymbol{x}) \in \mathbb{R}^+$, $R(\boldsymbol{x}) \in \textrm{Spin}(2) \cong \textrm{U}(1)$,
	it is just the even part of the rotated quantity $R^\dagger \Gamma R$, 
	averaged over space with weight $\rho^2$.
	
	It is instructive to check that e.g. vector-valued multiplication operators $\boldsymbol{v}$
	are hermitian under the inner product \eqref{inner_prod} since 
	\begin{equation} \label{}
		\langle (\boldsymbol{v}\Phi)^\dagger \Psi \rangle_+ 
		= \langle \Phi^\dagger \boldsymbol{v}^\dagger \Psi \rangle_+ 
		= \langle \Phi^\dagger (\boldsymbol{v}\Psi) \rangle_+,
	\end{equation}
	while $\hat{\imath}$ and the Dirac operator $\nabla$ are anti-hermitian as seen by
	\begin{equation} \label{}
	\begin{array}{rl}
		\langle (\Phi I)^\dagger \Psi \rangle_+
		&= \langle I^\dagger \Phi^\dagger \Psi \rangle_+ 
		 =-\langle I \Phi^\dagger \Psi \rangle_+
		 =-\langle I \Phi^\dagger \Psi \rangle_0 + \langle I \Phi^\dagger \Psi I \rangle_0 I \\
		&=-\langle \Phi^\dagger \Psi I \rangle_0 + \langle \Phi^\dagger \Psi I^2 \rangle_0 I
		 =-\langle \Phi^\dagger (\Psi I) \rangle_+,
	\end{array}
	\end{equation}
	and for smooth wavefunctions of compact support,
	$\Phi,\Psi \in C^{\infty}_0(\mathbb{R}^2; \mathcal{G})$,
	\begin{equation} \label{}
	\begin{array}{rl}
		\langle \nabla\Phi, \Psi \rangle 
		&= \sum_j \int \left\langle (\boldsymbol{e}_j \partial_j \Phi)^\dagger \Psi \right\rangle_+
		 = \sum_j \int \left\langle \partial_j(\Phi^\dagger) \boldsymbol{e}_j \Psi \right\rangle_+ \\
		&= \sum_j \int \partial_j \left( \langle \Phi^\dagger \boldsymbol{e}_j \Psi \rangle_+ \right) - \sum_j \int \left\langle \Phi^\dagger \boldsymbol{e}_j \partial_j(\Psi) \right\rangle_+ \\
		&= 0 - \langle \Phi, \nabla \Psi \rangle.
	\end{array}
	\end{equation}
	Therefore the anti-hermitian supercharge $\tilde{Q}$ really \emph{is} anti-hermitian
	w.r.t. the inner product
	and we have e.g. (as expected from supersymmetry, of course) that the expectation
	value of the identity in the state $\tilde{Q}\Psi$ is the expectation value of
	the hamiltonian (i.e. the energy) in the state $\Psi$,
	\begin{equation} \label{}
		\langle 1 \rangle_{\tilde{Q}\Psi} = \langle \tilde{Q}\Psi, \tilde{Q}\Psi \rangle 
		= - \langle \Psi, \tilde{Q}^2 \Psi \rangle  = \langle H \rangle_\Psi.
	\end{equation}
	
	We can also find a geometric interpretation of the operator $P$. 
	Defining the unit vector $\boldsymbol{n} := \frac{1}{\sqrt{2}} (\boldsymbol{e}_x - \boldsymbol{e}_y)$,
	we obtain from \eqref{gamma_p}
	\begin{equation} \label{}
		(P \Psi)(x,y) = \boldsymbol{n}\Psi(y,x).
	\end{equation}
	Note that the hyperplane orthogonal to $\boldsymbol{n}$ is the line of reflection $x=y$ in this case, 
	and we find that $P$ is (almost) a `square root of a reflection' in the sense that
	for multiplication operators $\Gamma\!: \mathbb{R}^2 \to \mathcal{G}(\mathbb{R}^2)$,
	\begin{equation} \label{reflection_exp_val}
		\langle \Gamma \rangle_{P\Psi} = \langle \bar{P}\Gamma^\star \rangle_{\Psi},
	\end{equation}
	where $(\bar{P} \Gamma)(x,y) := \boldsymbol{n} \Gamma(y,x)^\star \boldsymbol{n}$ is the 
	expected action of such a reflection on multivector fields inherited from 
	the action $\boldsymbol{n}\boldsymbol{v}(y,x)^\star\boldsymbol{n}$ on vector fields.
	Namely, using that
	\begin{equation} \label{}
		\big( (P\Psi)^\dagger \Gamma (P\Psi) \big)(x,y) = \Psi(y,x)^\dagger \boldsymbol{n}\Gamma(x,y)\boldsymbol{n} \Psi(y,x)
	\end{equation}
	and a change of variables $x \leftrightarrow y$ in the integral, we have
	\begin{equation} \label{}
	\begin{array}{rl}
		\langle \Gamma \rangle_{P\Psi}
		&= \int \langle (P\Psi)^\dagger \Gamma (P\Psi) \rangle_+
		 = \int \langle \Psi(y,x)^\dagger \boldsymbol{n} \Gamma(x,y) \boldsymbol{n} \Psi(y,x) \rangle_+ dxdy \\
		&= \int \langle \Psi(x,y)^\dagger \boldsymbol{n} \Gamma(y,x) \boldsymbol{n} \Psi(x,y) \rangle_+ dxdy
		 = \langle \bar{P} \Gamma^\star \rangle_\Psi.
	\end{array}
	\end{equation}
	As an example, consider the expectation value of the vector field $\Gamma = x\boldsymbol{e}_x$
	in the `reflected' state $P\Psi$,
	\begin{equation} \label{}
		\langle x\boldsymbol{e}_x \rangle_{P\Psi} 
		= \langle \bar{P}(-x\boldsymbol{e}_x) \rangle_{\Psi} 
		= \langle y\boldsymbol{n}\boldsymbol{e}_x\boldsymbol{n} \rangle_{\Psi} 
		= \langle -y\boldsymbol{e}_y \rangle_{\Psi}.
	\end{equation}
	Note that, because of the
	$\star$ in \eqref{reflection_exp_val}, this will
	yield a true reflection on even multivectors and minus a reflection on
	odd multivectors. It is however possible to define a true square root of a
	reflection by $(\tilde{P}\Psi)(x,y) := I\boldsymbol{n}\Psi(y,x)$ since this
	operator $\tilde{P}$ is also hermitian and 
	$(\tilde{P}\Gamma\tilde{P})(x,y) = -\boldsymbol{n}I\Gamma(y,x)I\boldsymbol{n} = \boldsymbol{n}\Gamma(y,x)^\star\boldsymbol{n}$.

\subsection{Coordinate transformation}

	It is interesting to note that $H_F = \nabla\frac{1}{2}(x^2 - y^2) = -\nabla(xy)I$.
	This suggests that we consider the coordinate transformation 
	\begin{equation} \label{coord_transf}
		\textstyle
		\boldsymbol{x} = (x,y) \quad \mapsto \quad \boldsymbol{u} = (u,v) := \left( \frac{1}{2}(x^2-y^2),xy \right).
	\end{equation}
	Note that this is exactly the complex transformation $z \mapsto w(z) := \frac{1}{2}z^2$.
	It is conformal everywhere except at the origin and maps the plane to
	itself twice. So, if we leave aside global issues of this transformation
	and instead focus on its local properties, then we can e.g. restrict attention to
	the open right half-plane $\mathbb{R}^2_+$ 
	which maps conformally and bijectively to the whole plane
	with the negative half-line removed.
	
	Consider first the bosonic hamiltonian $H_B$ acting on scalar wavefunctions. 
	Since the transformation is conformal, we have a simple transformation rule
	for the scalar laplacian,
	\begin{equation} \label{}
		\Delta_{xy} = |w'(z)|^2 \Delta_{uv} = h^{-2} \Delta_{uv},
	\end{equation}
	where $\Delta_{uv} = \partial_u^2 + \partial_v^2$ 
	and the scale factor $h$ can also be found from 
	\begin{equation} \label{}
		dx^2 + dy^2 = h^2(du^2 + dv^2),
	\end{equation}
	i.e. $h^{-1} := |\boldsymbol{x}| = \sqrt{2|\boldsymbol{u}|}$. 
	Using $dxdy = h^2 dudv$ for the integral measure we then have,
	for $\Phi,\Psi \in C^{\infty}_0(\mathbb{R}^2_+; \mathcal{G}^+)$,
	\begin{equation} \label{}
	\begin{array}{rl}
		\langle \Phi, H_B \Psi \rangle_{xy}
		&= \int \Phi^\dagger (x,y) (-\Delta_{xy} + x^2y^2) \Psi(x,y) dxdy \\
		&= \int (h\Phi)^\dagger (u,v) (-h^{-1}\Delta_{uv}h^{-1} + v^2) (h\Psi)(u,v) dudv \\
		&= \langle \tilde{\Phi}, \tilde{H}_B \tilde{\Psi} \rangle_{uv},
	\end{array}
	\end{equation}
	where we have introduced transformed wavefunctions $\tilde{\Psi} := h\Psi$
	and a transformed hamiltonian $\tilde{H}_B := -h^{-1}\Delta_{uv}h^{-1} + v^2$.
	In particular, the eigenvalue equation $H_B\Psi = \lambda\Psi$ becomes
	\begin{equation} \label{}
		0 = (H_B - \lambda)\Psi = h^{-1}(\tilde{H}_B - \lambda)\tilde{\Psi} 
		= h^{-2}\left( -\Delta_{uv} + \frac{v^2-\lambda}{2\sqrt{u^2+v^2}} \right)\Psi.
	\end{equation}
	
	Let us now turn to the case of wavefunctions with spin, i.e. we consider
	the transformation properties of the supersymmetry operators $Q$ and $H$.
	In order to ease the application of standard techniques,
	we will do much of this in the conventional complex representation. 
	However, because of the 1-to-1 correspondence
	\eqref{corresp_psi}-\eqref{corresp_i} between them, it is possible to translate each step
	into the geometric representation.
	First note that $Q$ involves a Dirac operator and the eigenvalue equation $Q\psi = \lambda\psi$
	has the form of a 2D Dirac equation, $(\gamma_1 \partial_x + \gamma_2 \partial_y + \Gamma)\psi = 0$,
	with $\Gamma \in C^{\infty}(\mathbb{R}^2; \mathbb{C}^{2 \times 2})$.
	Now, denoting $\boldsymbol{x} = (x_j)_{j=1,2}$, $\boldsymbol{u} = (\tilde{x}_j)_{j=1,2}$, 
	etc. as conventional, we have
	\begin{equation} \label{transformed_dirac}
		\textstyle
		\left( \sum_j \gamma_j \partial_j + \Gamma \right) \psi 
		= \left( h^{-1} \sum_j \tilde{\gamma}_j \tilde{\partial}_j + \Gamma \right) \psi,
	\end{equation}
	with space-dependent matrices 
	$\tilde{\gamma}_j := h \sum_k \frac{\partial \tilde{x}_j}{\partial x_k} \gamma_k$, i.e.
	$\tilde{\gamma}_1 = h( x\gamma_1 - y\gamma_2 )$ and
	$\tilde{\gamma}_2 = h( y\gamma_1 + x\gamma_2 )$.
	This amounts to introducing the space-dependent orthonormal basis $\{\boldsymbol{e}_u,\boldsymbol{e}_v\}$ 
	corresponding to the orthogonal coordinate system $(u,v)$, i.e.
	\begin{equation} \label{}
	\begin{array}{rl}
		\boldsymbol{e}_u &:= |\nabla u|^{-1} \nabla u = h (x \boldsymbol{e}_x - y \boldsymbol{e}_y), \\
		\boldsymbol{e}_v &:= |\nabla v|^{-1} \nabla v = h (y \boldsymbol{e}_x + x \boldsymbol{e}_y).
	\end{array}
	\end{equation}
	In the geometric representation we therefore have e.g.
	\begin{equation} \label{}
		\tilde{Q} = h^{-1}(\boldsymbol{e}_u \partial_u + \boldsymbol{e}_v \partial_v) + vI
	\end{equation}
	and
	\begin{equation} \label{}
		H = -h^{-2}\Delta_{uv} + v^2 + h^{-1}\boldsymbol{e}_u.
	\end{equation}
	It is conventional to transform away the space-dependence of the gamma matrices by
	applying a pointwise change of spinor basis, i.e. a local $\textrm{Spin}(2)$ gauge transformation
	$\psi \mapsto R\psi$.
	For this we make the observation that if we
	let $\phi := \arctan \frac{y}{x}$ denote the polar angle in the $(x,y)$-coordinate system,
	then
	\begin{equation} \label{}
	\begin{array}{rl}
		\boldsymbol{e}_u &= \cos{\phi}\ \boldsymbol{e}_x - \sin{\phi}\ \boldsymbol{e}_y = R^\dagger \boldsymbol{e}_x R, \\
		\boldsymbol{e}_v &= \sin{\phi}\ \boldsymbol{e}_x + \cos{\phi}\ \boldsymbol{e}_y = R^\dagger \boldsymbol{e}_y R = \boldsymbol{e}_u I,
	\end{array}
	\end{equation}
	with $R := e^{-\frac{\phi}{2}I}$.
	Hence, $\tilde{\gamma}_j = R^\dagger \gamma_j R$
	(we identify $I = \gamma_1\gamma_2$ according to the correspondence), 
	and \eqref{transformed_dirac} becomes
	\begin{equation} \label{transformed_dirac_cont}
		\textstyle
		\left( h^{-1} \sum_j R^\dagger \gamma_j R \tilde{\partial}_j + \Gamma \right) R^\dagger R \psi 
		= R^\dagger \left( h^{-1} \sum_j \gamma_j (\tilde{\partial}_j + R \tilde{\partial}_j R^\dagger) + R \Gamma R^\dagger \right) (R\psi), 
	\end{equation}
	so the cost of making the local gauge transformation is, naturally, 
	the addition of a non-vanishing connection term 
	$\Omega_j := R \tilde{\partial}_j R^\dagger = \frac{1}{2} \tilde{\partial}_j \phi I$.
	
	As described in \cite{schluter_et_al}, it is
	always possible to `absorb' the connection resulting from a conformal
	transformation by rescaling 
	(alt. see Theorem 5.24, Chapter II in \cite{lawson_michelsohn}).
	Namely, observing that
	\begin{equation} \label{}
		\partial_u \phi = -\partial_v \ln{h^{-1}}, \quad
		\partial_v \phi =  \partial_u \ln{h^{-1}},
	\end{equation}
	we find
	\begin{equation} \label{conn_vs_scaling}
		\textstyle
		\sum_j \gamma_j (R \tilde{\partial}_j R^\dagger + h^{\frac{1}{2}} \tilde{\partial}_j h^{-\frac{1}{2}}) = 0,
	\end{equation}
	and hence \eqref{transformed_dirac} and \eqref{transformed_dirac_cont} give
	\begin{equation} \label{}
		\textstyle
		\left( \sum_j \gamma_j \partial_j + \Gamma \right) \psi
		= h^{-\frac{1}{2}} R^\dagger \left( h^{-1} \sum_j \gamma_j \tilde{\partial}_j + R \Gamma R^\dagger \right) (h^{\frac{1}{2}}R\psi). 
	\end{equation}
	Note that we can think of the constant matrices $\gamma_j$ as now
	representing the basis vectors $\boldsymbol{e}_u, \boldsymbol{e}_v$
	(but in a transformed frame where these are now constant), since e.g.
	$\Gamma = \tilde{\gamma}_1 \leftrightarrow \boldsymbol{e}_u$ transforms into
	$R \Gamma R^\dagger = \gamma_1$. 
	Denoting $\tilde{\boldsymbol{e}}_u := \boldsymbol{e}_x \leftrightarrow \gamma_1$,
	$\tilde{\boldsymbol{e}}_v := \boldsymbol{e}_y \leftrightarrow \gamma_2$,
	and $\nabla_{\!uv} := \tilde{\boldsymbol{e}}_u \partial_u + \tilde{\boldsymbol{e}}_v \partial_v$,
	we find that the eigenvalue equation $(Q - \lambda)\Psi = 0$ can be written
	\begin{equation} \label{}
		\left( \nabla_{\!uv} + \frac{ vI - \lambda\hat{\imath} }{ \sqrt{2}(u^2 + v^2)^{\frac{1}{4}} } \right)(h^{\frac{1}{2}}R\Psi) = 0.
	\end{equation}
	With respect to inner products we have, 
	for $\Phi, \Psi \in C_0^\infty(\mathbb{R}^2_+; \mathcal{G})$,
	\begin{equation} \label{}
	\begin{array}{rl}
		\langle \Phi, \tilde{Q} \Psi \rangle_{xy}
		&= \int \left\langle \Phi^\dagger (x,y) (\nabla_{xy} + xyI) \Psi(x,y) \right\rangle_+ dxdy \\
		&= \int \left\langle (h^{\frac{1}{2}}R\Phi)^\dagger (u,v) (\nabla_{\!uv} + hvI) (h^{\frac{1}{2}}R\Psi)(u,v) \right\rangle_+ dudv \\
		&= \langle \hat{\Phi}, \hat{Q} \hat{\Psi} \rangle_{uv},
	\end{array}
	\end{equation}
	with transformed wavefunctions $\hat{\Psi} := hR\Psi$ and anti-hermitian supercharge 
	\begin{equation} \label{}
		\hat{Q} := h^{-\frac{1}{2}} \nabla_{\!uv} h^{-\frac{1}{2}} + vI.
	\end{equation}
	The square of this supercharge,
	\begin{equation} \label{}
		\hat{Q}^2 = h^{-\frac{1}{2}} \nabla_{\!uv} h^{-1} \nabla_{\!uv} h^{-\frac{1}{2}} 
		+ h^{-\frac{1}{2}} \left(\dot{\nabla}_{\!uv}\dot{v}I + v\nabla_{\!uv}I + vI\nabla_{\!uv}\right) h^{-\frac{1}{2}} 
		- v^2,
	\end{equation}
	yields minus the expected transformed hamiltonian
	\begin{equation} \label{}
		\hat{H} := -h^{-1}\hat{\Delta}h^{-1} + v^2 + h^{-1}\tilde{\boldsymbol{e}}_u,
	\end{equation}
	where $\hat{\Delta}$ is the connection laplacian (see \cite{lawson_michelsohn}).
	Namely, using \eqref{conn_vs_scaling} we have
	\begin{equation} \label{}
	\begin{array}{rl}
		\hat{\Delta} &:= h^{\frac{1}{2}} \nabla_{\!uv} h^{-\frac{1}{2}} h^{-\frac{1}{2}} \nabla_{\!uv} h^{\frac{1}{2}}
		= (\nabla_{\!uv} + h^{\frac{1}{2}} \dot{\nabla}_{\!uv} \dot{h}^{-\frac{1}{2}}) (\nabla_{\!uv} - h^{\frac{1}{2}} \dot{\nabla}_{\!uv} \dot{h}^{-\frac{1}{2}}) \\
		&= \sum_{j,k} \tilde{\boldsymbol{e}}_j(\tilde{\partial}_j - \Omega_j) \tilde{\boldsymbol{e}}_k(\tilde{\partial}_k + \Omega_k)
		= \sum_j (\tilde{\partial}_j + \Omega_j)^2.
	\end{array}
	\end{equation}
	Furthermore, using that $\nabla_{\!uv} \ln{h^{-\frac{1}{2}}} = h^4 \boldsymbol{u}$,
	with $\boldsymbol{u} = u\tilde{\boldsymbol{e}}_u +  v\tilde{\boldsymbol{e}}_v$,
	and that $\Delta_{uv} \ln{h^{-\frac{1}{2}}} = \frac{1}{4} \Delta_{uv} \textrm{Re} \log{w} = 0$, we obtain
	\begin{equation} \label{}
	\begin{array}{rl}
		\hat{\Delta} &= (\nabla_{\!uv} + \dot{\nabla}_{\!uv} \ln{\dot{h}^{-\frac{1}{2}}}) (\nabla_{\!uv} - \dot{\nabla}_{\!uv} \ln{\dot{h}^{-\frac{1}{2}}})
		= (\nabla_{\!uv} + h^4\boldsymbol{u}) (\nabla_{\!uv} - h^4\boldsymbol{u}) \\
		&= \Delta_{uv} - \dot{\nabla}_{\!uv} (h^4\boldsymbol{u})\dot{} - h^4\dot{\nabla}_{\!uv}\boldsymbol{u}(\ )\dot{} + h^4\boldsymbol{u}\nabla_{\!uv} - h^8\boldsymbol{u}^2 \\
		&= \Delta_{uv} - \frac{1}{4} h^4 + 2h^4\boldsymbol{u} \wedge \nabla_{\!uv}.
	\end{array}
	\end{equation}
	The last term contains an angular momentum operator 
	$\boldsymbol{u} \wedge \nabla_{\!uv} = I(u\partial_v - v\partial_u)$.
	Summing up, we have
	\begin{equation} \label{}
		h\hat{H}h = -\Delta_{uv} + \frac{v^2}{2|\boldsymbol{u}|} + \frac{1}{\sqrt{2|\boldsymbol{u}|}} \tilde{\boldsymbol{e}}_u + \frac{1}{16|\boldsymbol{u}|^2} (1 - 8\boldsymbol{u} \wedge \nabla_{\!uv}),
	\end{equation}
	which would diagonalize on the subspaces $(1 \pm \tilde{\boldsymbol{e}}_u) \mathscr{H}$,
	had it not been for the angular momentum term.
	Taking $u \to +\infty$ while keeping $v$ finite we recover, in terms of these coordinates,
	the observation of \cite{deWit_luscher_nicolai} that the negative energy of
	the fermionic part precisely cancels the minimal energy of the
	resulting harmonic oscillator in the $v$ coordinate,
	while remaining terms approach a free laplacian in $u$.
	With $u$ finite and $v \to \infty$ we (instead of an oscillator) 
	approach an Airy equation in $v$.
	
\subsection{Generalized model}
	
	Returning to the original cartesian coordinate system, 
	we have noted that $H = -\Delta + v^2 - (\nabla v)I$
	with $v = xy$.
	It is straightforward to generalize this setup.
	In particular, we can for any sufficiently regular scalar
	field $\varphi\!: \mathbb{R}^2 \to \mathbb{R}$ define the operator
	\begin{equation} \label{q_phi}
		\textstyle
		\tilde{Q}_\varphi := \nabla + \varphi I
	\end{equation}
	and find a corresponding `supersymmetric hamiltonian'
	\begin{equation} \label{h_phi}
		H_\varphi := -\tilde{Q}_\varphi^2 
		= -(\nabla^2 + \varphi^2 I^2 + \dot{\nabla}\dot{\varphi}I + \varphi \nabla I + \varphi I \nabla)
		= -\Delta + \varphi^2 - (\nabla \varphi)I.
	\end{equation}
	Furthermore, since the simplification in the last step 
	only relies on the fact that $I$ anticommutes with vectors and squares to $-1$, 
	this operator algebra actually extends to every other even dimension $d=2,6,10,\ldots$.
	This coincides with the dimensions for which we have the grading operator $K$
	and the complex structure $\hat{\imath}$. Following the procedure above for the
	choice of a hermitian inner product on the Hilbert space 
	$\mathscr{H}^{(d)} := L^2(\mathbb{R}^d; \mathcal{G}(\mathbb{R}^d))$
	using this given complex structure, we find that it can be written
	\begin{equation} \label{general_scalar_prod}
		\langle \Phi, \Psi \rangle := \int_{\mathbb{R}^d} \langle \Phi^\dagger \Psi \rangle_{0,d}.
	\end{equation}
	This inner product takes values in the subspace 
	$\langle \mathcal{G} \rangle_0 \oplus \langle \mathcal{G} \rangle_d \cong \mathbb{C}$
	and hermiticity of operators with respect to it, etc. follows exactly as above.

	Given some unit vector $\boldsymbol{n} \in \mathbb{R}^d$, it is natural to define
	corresponding generalized reflection operators 
	$P_{\boldsymbol{n}}$, $\tilde{P}_{\boldsymbol{n}}$, and $\bar{P}_{\boldsymbol{n}}$ by
	\begin{equation} \label{}
		(P_{\boldsymbol{n}} \Psi)(\boldsymbol{x}) := \boldsymbol{n} \Psi(-\boldsymbol{nxn}), \quad
		(\tilde{P}_{\boldsymbol{n}} \Psi)(\boldsymbol{x}) := I\boldsymbol{n} \Psi(-\boldsymbol{nxn}),
	\end{equation}
	and
	\begin{equation} \label{}
		(\bar{P}_{\boldsymbol{n}} \Gamma)(\boldsymbol{x}) := \boldsymbol{n} \Gamma(-\boldsymbol{nxn})^\star \boldsymbol{n}
		= (P_{\boldsymbol{n}} \Gamma^\star P_{\boldsymbol{n}})(\boldsymbol{x}) = (\tilde{P}_{\boldsymbol{n}} \Gamma \tilde{P}_{\boldsymbol{n}})(\boldsymbol{x}),
	\end{equation}
	for $\Psi \in \mathscr{H}^{(d)}$ and multivector fields $\Gamma\!: \mathbb{R}^d \to \mathcal{G}(\mathbb{R}^d)$.
	It is easy to verify that $\{P_{\boldsymbol{n}},\tilde{Q}_\varphi\} = 0$ whenever $\boldsymbol{n}$ is
	a direction of reflection symmetry of $\varphi$, i.e. whenever
	$\varphi(-\boldsymbol{nxn}) = \varphi(\boldsymbol{x})$,
	while $[\tilde{P}_{\boldsymbol{n}},\tilde{Q}_\varphi] = 0$ for an antisymmetry,
	$\varphi(-\boldsymbol{nxn}) =-\varphi(\boldsymbol{x})$.
	Hence, even in this general setting, as long as $\varphi$ possesses such a
	reflection symmetry we obtain a supersymmetric system
	$(\mathscr{H}^{(d)}, H_\varphi, P_{\boldsymbol{n}}, Q_\varphi)$
	with hermitian supercharge 
	$Q_\varphi := -\hat{\imath}\tilde{Q}_\varphi = -\hat{\imath}\nabla + \varphi K$.

	As an application of this 
	general framework
	we can consider a
	higher-dimen\-sional analogue of the toy model. We take as a scalar field
	(with a number of reflection symmetries)
	\begin{equation} \label{}
		\varphi(x_1,x_2,\ldots,x_d) := x_1 x_2 \ldots x_d
	\end{equation}
	so that
	\begin{equation} \label{}
		\tilde{Q}_\varphi = \nabla + x_1 x_2 \ldots x_d I
	\end{equation}
	and
	\begin{equation} \label{}
		H_\varphi = -\Delta + x_1^2 x_2^2 \ldots x_d^2 - \sum_{k=1}^d x_1 \ldots x_{k-1} x_{k+1} \ldots x_d \boldsymbol{e}_k I.
	\end{equation}
	Just as in the case $d=2$ for the toy model, the bosonic part of this hamiltonian,
	$H_B = -\Delta + x_1^2 \ldots x_d^2$, has a strictly positive and
	purely discrete spectrum (use e.g. Prop. 6 in \cite{garcia_del_moral_et_al}).
	Furthermore, the fermionic part $H_F = -(\nabla\varphi)I$ 
	is still just a multiplication operator and satisfies
	\begin{equation} \label{}
		H_F^2 = (\nabla\varphi)I (\nabla\varphi)I = -(\nabla\varphi)^2 I^2 = (\nabla\varphi)^2
		= \sum_{k=1}^d x_1^2 \ldots x_{k-1}^2 x_{k+1}^2 \ldots x_d^2.
	\end{equation}
	Hence, it has in each point eigenvalues 
	\begin{equation} \label{}
		\pm |\nabla\varphi| = \pm \sqrt{\sum_{k=1}^d x_1^2 \ldots x_{k-1}^2 x_{k+1}^2 \ldots x_d^2}
	\end{equation}
	on the spinor subspaces (ideals) 
	$\left( 1 \mp \frac{\nabla\varphi}{|\nabla\varphi|} I \right) \mathcal{G}(\mathbb{R}^d)$, respectively.
	Note that if e.g. $x_d$ is taken very small compared to $x_1,\ldots,x_{d-1}$ then
	these eigenvalues tend to $\pm |x_1 x_2 \ldots x_{d-1}|$.
	This suggests the following asymptotic analysis: 
	Let, say, $x_1 \sim x_2 \sim \ldots \sim x_{d-1} \sim \chi$ (slow variables)
	and take $\chi \to \infty$ while keeping the coordinate $x_d$ finite (fast variable).
	Then
	\begin{equation} \label{}
		H_F \ \sim\ \sum_{k=1}^{d-1} \chi^{d-2} x_d \boldsymbol{e}_kI + \chi^{d-1} \boldsymbol{e}_dI \ \sim\ \chi^{d-1} \boldsymbol{e}_dI
	\end{equation}
	and
	\begin{equation} \label{}
		H_B \ \sim\ -\sum_{k=1}^{d-1} \partial_k^2 - \partial_d^2 + \chi^{2(d-1)} x_d^2 
		\ \sim\ -\sum_{k=1}^{d-1} \partial_k^2 + \chi^{d-1} \left( -\frac{1}{\chi^{d-1}} \partial_d^2 + \chi^{d-1} x_d^2 \right),
	\end{equation}
	again allowing for a cancellation of minimal energies,
	which suggests that the spectrum of $H_\varphi$ is continuous.

	Another SUSYQM is obtained by taking $\varphi(\boldsymbol{x}) := |\boldsymbol{x}|$, i.e.
	\begin{equation} \label{}
		\tilde{Q}_{\varphi} = \nabla + |\boldsymbol{x}|I
	\end{equation}
	and
	\begin{equation} \label{}
		H_{\varphi} = -\Delta + \boldsymbol{x}^2 - \boldsymbol{e}_rI,
	\end{equation}
	where $\boldsymbol{e}_r := \boldsymbol{x}/|\boldsymbol{x}| = \nabla |\boldsymbol{x}|$.
	In this case, $H_B = -\Delta + \boldsymbol{x}^2$ is an ordinary $d$-dimensional 
	harmonic oscillator and $H_F = -\boldsymbol{e}_rI$ has pointwise eigenspaces
	$(1 \mp \boldsymbol{e}_rI)\mathcal{G}$ with eigenvalue $\pm 1$.
	Taking expectation values, we find $H_{\varphi} \geq d - 1 \gg 0$, 
	and hence this system does not possess a supersymmetric ground state.

\section{An alternative higher-dimensional supersymmetric harmonic oscillator}

	Witten \cite{witten_81,witten_82}, de Crombrugghe and Rittenberg \cite{crombrugghe_rittenberg}, 
	and many others have considered the simple $\mathcal{N} = 2$
	SUSYQM system described by the supercharges
	\begin{equation} \label{q_witten}
	\begin{array}{rl}
		Q_1 &= p_x \sigma_1 + W(x) \sigma_2, \\
		Q_2 &= p_x \sigma_2 - W(x) \sigma_1,
	\end{array}
	\end{equation}
	satisfying
	\begin{equation} \label{q_witten_squared}
		Q_1^2 \ = \ Q_2^2 \ = \ p_x^2 + W(x)^2 + W'(x) \sigma_3.
	\end{equation}
	As a special case, $W(x) := x$, one has
	\begin{equation} \label{one_dim_osc}
		H_x := Q_1^2 = Q_2^2 = p_x^2 + x^2 + \sigma_3,
	\end{equation}
	i.e. a one-dimensional supersymmetric harmonic oscillator \cite{nicolai}.
	
	Operators of the form \eqref{q_witten} arise e.g. when studying the
	Dirac equation in 1+1 spacetime dimensions.
	However, we would like to investigate whether there is an alternative geometric
	interpretation underlying the presence of the two Clifford generators $\sigma_1$ and $\sigma_2$.
	These pair up to form the fermionic oscillator term $\sigma_3 = -i\sigma_1\sigma_2$
	(alternatively, in the particle interpretation, 
	they pair up to form fermionic creation and annihilation operators
	$c^\dagger$ and $c$ s.t. $\frac{1}{2}\sigma_3 = c^\dagger c - \frac{1}{2}$).
	Along the lines of our approach to previous examples, we choose to consider the Clifford
	generators as being more fundamental and generating the real
	geometric algebra $\mathcal{G}(\mathbb{R}^2)$, 
	with $\sigma_1 \leftrightarrow \boldsymbol{e}_x$ and $\sigma_2 \leftrightarrow \boldsymbol{e}_y$.
	Moreover, we consider the extension of the bosonic coordinate space $\mathbb{R}$ to $\mathbb{R}^2$
	and let the supercharges \eqref{q_witten} act on different coordinates $x$ and $y$,
	while still `sharing the same fermion' by involving the 
	same pair of Clifford generators $\boldsymbol{e}_x$ and $\boldsymbol{e}_y$.
	Thus, for the case of the oscillator potential, we define
	\begin{equation} \label{q_x_y}
	\begin{array}{rl}
		Q_x &:= p_x \boldsymbol{e}_x + x \boldsymbol{e}_y, \\
		Q_y &:= p_y \boldsymbol{e}_y - y \boldsymbol{e}_x.
	\end{array}
	\end{equation}
	Separately, we have a pair of supersymmetric oscillators,
	\begin{equation} \label{}
		Q_x^2 = p_x^2 + x^2 + K, \qquad
		Q_y^2 = p_y^2 + y^2 + K,
	\end{equation}
	(where as usual $K := -\hat{\imath}\boldsymbol{e}_x\boldsymbol{e}_y$ in 2D),
	but combined, $Q := Q_x + Q_y$, we find
	\begin{equation} \label{}
		Q^2 = Q_x^2 + Q_y^2 + \{Q_x,Q_y\} = p_x^2 + p_y^2 + x^2 + y^2 + 2K + xp_y - yp_x.
	\end{equation}
	This is a two-dimensional bosonic oscillator together with 
	two copies of the same fermionic oscillator, 
	plus the angular momentum operator in two dimensions
	which gives is an indication of the rotational symmetry now present in the pair
	of supercharges \eqref{q_x_y}.
	Note that we can write
	\begin{equation} \label{q_harmonic_2d}
		Q = p_x\boldsymbol{e}_x + p_y\boldsymbol{e}_y + (x\boldsymbol{e}_x + y\boldsymbol{e}_y) \boldsymbol{e}_x\boldsymbol{e}_y = -\hat{\imath}\nabla + \boldsymbol{x}I.
	\end{equation}
	As in our previous examples, 
	the supercharge \eqref{q_harmonic_2d} generalizes straightforwardly to the dimensions 
	where the algebra works out. In particular, on $\mathscr{H}^{(d)}$ we define
	\begin{equation} \label{q_harmonic_d}
		Q := -\hat{\imath}\nabla + \boldsymbol{x}I,
	\end{equation}
	and for $d=2,6,10,\ldots$ we again see a simplification in
	the expression for the square of this operator:
	\begin{equation} \label{}
	\begin{array}{rl}
		Q^2 &= -\nabla^2 + \boldsymbol{x}I\boldsymbol{x}I - \hat{\imath} \left( \dot{\nabla}\dot{\boldsymbol{x}} I + \sum_j \boldsymbol{e}_j \boldsymbol{x}I \partial_j + \sum_j \boldsymbol{x}I \boldsymbol{e}_j \partial_j \right) \\
		&= -\Delta + \boldsymbol{x}^2 + K(d - 2L).
	\end{array}
	\end{equation}
	Here,
	\begin{equation} \label{}
		\textstyle
		L := \boldsymbol{x} \wedge \nabla = \sum_{j<k} \boldsymbol{e}_j\boldsymbol{e}_k (x_j\partial_k - x_k\partial_j)
	\end{equation}
	is a generalized angular momentum operator in the sense that each
	bivector component $\boldsymbol{e}_j \wedge \boldsymbol{e}_k$ contains the corresponding
	generator $\hat{\imath}L_{jk}$ of rotations in the $x_j x_k$-plane.
	In particular, $L\Psi = 0$ if $\Psi$ is rotationally invariant.
	
	The above observations suggest the following interesting interpretation of this system.
	Consider the hamiltonian
	\begin{equation} \label{h_alt_susy_osc}
		H := -\Delta + \boldsymbol{x}^2 + Kd.
	\end{equation}
	This can be viewed as a sum of hamiltonians of the form $H_x$
	for each coordinate, hence the $d$-dimensional bosonic oscillator term
	$H_B = -\Delta + \boldsymbol{x}^2$, but each is accompanied
	by the same fermionic term $K$, resulting in $H_F = Kd$.
	We have that $Q$ anticommutes with $K$, and $K^2 = 1$,
	but with respect to the hamiltonian $H$ we do not quite have a SUSYQM 
	according to the definition since
	\begin{equation} \label{q_squared_constraint}
		Q^2 = H - 2KL.
	\end{equation}
	However, note that $H$ is rotationally symmetric. If we consider the subspace 
	of rotationally symmetric states, 
	$\mathscr{H}_\textrm{sym} := \textrm{ker}\ \langle L^2 \rangle_0 = \textrm{ker} \sum_{j<k} L_{jk}^2$, 
	we have there the supersymmetry algebra $Q^2 = H$. 
	Furthermore, we know that the spectrum of $H$ is $\{2n_1 + \ldots + 2n_d + d \pm d\}_{n_j = 0,1,2,\ldots}$
	and zero energy states are explicitly given by 
	$\Psi_0(\boldsymbol{x}) = e^{-\frac{1}{2}\boldsymbol{x}^2} \Psi_- \in \mathscr{H}^{(d)}_-$,
	where $\Psi_- \in \mathcal{G}^-$ is a constant odd spinor.
	These \emph{do} lie in $\mathscr{H}_\textrm{sym}$ and are thus supersymmetric.

	It is also tempting to view the resulting system \eqref{q_squared_constraint}
	as a SUSYQM with constraints.
	Namely, say that we start with the kinematical Hilbert space $\mathscr{H}^{(d)}$ 
	and the hamiltonian $H$ with rotational invariance as a gauge symmetry,
	expressed by $[H,L] = 0$.
	Let us require \emph{physical} states to be gauge-invariant, 
	hence we consider the subspace of states annihilated by the constraint operator $L$. 
	On this physical Hilbert space of gauge-invariant states, 
	$\mathscr{H}_\textrm{phys} := \textrm{ker}\ L$,
	we have a `supersymmetric system' $(\mathscr{H}_\textrm{phys},H,K,Q)$.
	However, this is not a SUSYQM in the strict sense since $[Q,L] \neq 0$
	and $\mathscr{H}_\textrm{phys}$ is not invariant under $Q$.
	
	As in previous examples, we can lighten our conditions a bit and
	instead of $Q$ consider the anti-hermitian
	supercharge $\tilde{Q} := \hat{\imath}Q = \nabla - \boldsymbol{x}K$.
	We have $\{\tilde{Q},K\} = 0$ and
	\begin{equation} \label{}
		\tilde{Q}^2 = \Delta + \boldsymbol{x}K\boldsymbol{x}K - \dot{\nabla}\dot{\boldsymbol{x}}K - \dot{\nabla}\boldsymbol{x}K(\ )\dot{} - \boldsymbol{x}K\nabla
		= - (H - 2KL).
	\end{equation}
	We can even take as $K$ the grade involution operator defined
	in all dimensions, and since we have removed the need for a
	complex structure we find that we have this type of
	`supersymmetric system' in arbitrary dimensions.
	Furthermore, it is also possible to define
	\begin{equation} \label{q_vector_field}
		\tilde{Q} := \nabla - \boldsymbol{f}K
	\end{equation}
	for any sufficiently regular vector field $\boldsymbol{f}\!: \mathbb{R}^d \to \mathbb{R}^d$ 
	and in that case obtain
	\begin{equation} \label{h_vector_field}
		-\tilde{Q}^2 = -\Delta + |\boldsymbol{f}|^2 + K \big( \dot{\nabla} \dot{\boldsymbol{f}} - 2\boldsymbol{f} \wedge \nabla \big),
	\end{equation}
	with $\nabla \boldsymbol{f} = \textrm{div} \boldsymbol{f} + \nabla \wedge \boldsymbol{f}$.

	A system similar to \eqref{h_alt_susy_osc}-\eqref{q_squared_constraint}
	above has been considered in \cite{delbourgo}, but using parity instead
	of spinor grades, and only in $d=4$ for the explicitly supersymmetric case.
	See also e.g. \cite{plyushchay}, where the parity operator is used to
	add supersymmetry to otherwise purely bosonic systems.

\section{Conclusion}

	By reformulating the supersymmetric systems we started out from in terms of
	geometric algebra, we could easily identify them as special cases
	of the following:
	\begin{displaymath} \label{}
	\begin{array}{lll}
		\textrm{Dirac operator (with magnetic field):} \\
		H = \sum_j (-\hat{\imath}\partial_j - A_j)^2 + \hat{\imath} \nabla \wedge \boldsymbol{A}, \quad &
		K, \quad &
		Q = -\hat{\imath}\nabla - \boldsymbol{A} \\
		\textrm{Supermembrane toy model:} \\
		H = -\Delta + \varphi^2 - (\nabla \varphi)I, \quad &
		P_{\boldsymbol{n}}, \quad &
		Q = -\hat{\imath}\nabla + \varphi K \\
		\textrm{Oscillator:} \\
		H = -\Delta + |\boldsymbol{f}|^2 + \big( \textrm{div} \boldsymbol{f} + \nabla \wedge \boldsymbol{f} - 2\boldsymbol{f} \wedge \nabla \big)K, \quad &
		K, \quad &
		Q = -\hat{\imath}\nabla + \boldsymbol{f}I
	\end{array}
	\end{displaymath}
	Not only do we have a clear geometric interpretation of every
	constituent of these systems, but relations between them are
	also simple to derive in this language.
	It is also interesting to note the similarities between these
	higher-dimensional
	systems and the well-studied one-dimensional toy model \eqref{q_witten_squared}.
	Furthermore, by relaxing the requirement of a canonical complex structure and
	hermiticity of the supercharges, we have seen that we can also find purely real
	analogues of these systems in arbitrary dimensions.

\newpage	

\section*{Acknowledgements}

	I thank Jens Hoppe and Lars Svensson for useful discussions 
	and valuable comments on the manuscript.
	I would also like to thank Volker Bach and Hubert Kalf for discussions
	and hospitality at Mainz University and King's College London, respectively,
	as well as Gian Michele Graf for discussions.

\end{document}